\documentclass[10pt]{article}
\usepackage{amsmath}
\usepackage{amsfonts}
\usepackage{graphicx}
\usepackage{authblk}
\usepackage{xcolor}
\usepackage{fullpage} 
\usepackage[labelfont=bf]{caption}
\newcommand{\etal}{\mbox{\emph{et al.\ }}}

\title{\bf Elastowetting of Soft Hydrogel Spheres}
\author[,1]{Aditi Chakrabarti\thanks{Present address: Paulson School of Engineering and Applied Sciences, Harvard University, Cambridge, Massachusetts 02138, United States.}}
\author[2]{Amir Porat}
\author[2]{Elie Rapha\"{e}l}
\author[3,4]{Thomas Salez}
\author[,1]{Manoj K. Chaudhury\thanks{Corresponding author: \texttt{mkc4@lehigh.edu}}}
\affil[1]{Department of Chemical and Biomolecular Engineering, Lehigh University, Bethlehem, Pennsylvania 18015, United States.}
\affil[2]{Laboratoire de Physico-Chimie Th\'{e}orique, UMR CNRS Gulliver 7083, ESPCI Paris, PSL Research University, Paris, France.}
\affil[3]{Univ. Bordeaux, CNRS, LOMA, UMR 5798, F-33405 Talence, France.}
\affil[4]{Global Station for Soft Matter, Global Institution for Collaborative Research and Education, Hokkaido University, Sapporo, Japan.}

\date{\today}
\begin{document}
\maketitle

\begin{abstract}
When a soft hydrogel sphere is placed on a rigid hydrophilic substrate, it undergoes arrested spreading by forming an axisymmetric foot near the contact line, while conserving its global spherical shape. In contrast, liquid water (that constitutes greater than 90\% of the hydrogel's volume) spreads into a thin film on the same surface. We study systematically this elastowetting of gel spheres on substrates of different surface energies, and find that their contact angle increases as the work of adhesion between the gel and the substrate decreases, as one would observe for drops of pure water -- albeit being larger than in the latter case. This difference in the contact angles of gel and water appears to be due to the elastic shear stresses that develop in the gel and oppose its spreading. Indeed, by increasing the elastic modulus of the gel spheres, we find that their contact angle also increases. In addition, the length of the contact €œfoot€ increases with the work of adhesion and sphere size, while it decreases when the elastic modulus of the gel is increased. We discuss those experimental results in light of a minimal analysis based on energy minimization, volume conservation, and scaling arguments.
\end{abstract}

\section{Introduction}
Contact mechanics is an important field that involves the study of how elastic solids deform when they are brought into contact with each other. While the Hertz model~\cite{Hertz1881} forms the basis to characterize infinitesimal deformations of non-adhesive stiff spheres, Johnson-Kendall-Roberts' (JKR) theory~\cite{Johnson1971,Johnson1985} describes situations in which adhesion becomes important, \textit{e.g.} for soft objects such as elastomers~\cite{chaudhury1991direct} and cells~\cite{Chu2005}. However, there is yet another regime involving much softer materials, such as hydrogels, where surface tension also contributes to their mechanics~\cite{Mora_prl2010}, thus implying that their contact behavior should fundamentally differ from that of stiffer materials. In this regime, we speculate that a soft elastic particle, when contacted with a rigid hydrophilic substrate, would undergo a large deformation with effects of both surface tension and elasticity. This so-called elastowetting of soft hydrogel spheres on rigid substrates can be considered to be an elastic perturbation to the classical Young€™-Dupr\'e wetting of drops of liquid~\cite{Degennes1985}. The contact mechanism of soft particles is expected to depend on hysteresis, phase separation~\cite{Jensen2015,Liu2016}, and specific material properties~\cite{Rivetti2017}. By studying the wetting behaviour of such soft materials, we expect to gain a better understanding of cell migration~\cite{Lo2000,Joanny2003,Schwarz2013,Fischer-Friedrich2014}, nanoparticle adhesion~\cite{champion2006role}, mechanical properties of nanomaterials~\cite{Cuenot2004}, behavior of pressure-sensitive adhesives~\cite{newby1995science}, and it may have important implications in the design of new approaches for atomic-level characterization~\cite{Luan2005,Erath2010}. 

The specific geometry of a rigid sphere contacting a soft elastic substrate, in presence of adhesion and solid-air surface tension, has been studied in detail recently, both experimentally and theoretically~\cite{rimai1989adhesion,rimai1994adhesion,Rimai2000,Style2013,Chakrabarti2014,Hui2015,Butt2016,Karpitschka2016,andreotti2016soft,ina2017adhesion}. These works assume geometrically small elastic deformations, which is only valid for a sufficiently low work of adhesion. Finite-element simulations allowed to extend this framework to larger deformations through a neo-Hookean approach~\cite{Xu2014}. In contrast, the dual geometry of a very soft elastic sphere placed atop a flat rigid substrate has only been scarcely addressed, despite its obvious similarity with the natural configuration of liquid droplet wetting and cell adhesion. On the experimental side, the spreading of latex particles on rigid substrates was studied using atomic force microscopy (AFM)~\cite{Lau2002}. However, there are two limitations in this preliminary work. First, due to the very small particle size ($\sim 0.1 \,\mu$m), it was only possible to image them in plan view. Therefore, instead of directly measuring the vertical deformation, the contact radius of the particles was used to estimate it -- based on an assumed spherical-cap shape€. Secondly, latex is a glassy material that can flow if the contact stresses exceed its yield point. On the theoretical side, there are currently only global scalings to describe the elastowetting of soft spheres~\cite{Carrillo2010,Salez2013}, which again assume a spherical-cap shape. What we expect in reality is a scenario where the adhesion-driven wetting of the elastic sphere gives rise to a gradient of shear deformation, from the highest value closest to the contact region and decaying with the distance from the substrate. Finally, we note that the problem of spreading of a semi-cylindrical polymer gel placed on a flat rigid substrate was studied theoretically~\cite{Joanny2001}, in which an equilibrium shape was envisioned with a foot due to a positive spreading parameter $S$. The foot length $\ell$ was further predicted to be independent of the cylinder radius, and to scale only with the ratio of the spreading parameter and the shear modulus $\mu$, as $\ell \sim S/\mu$. We note that a similar foot-like feature was also observed in the dual geometry of a rigid sphere in contact with a soft substrate~\cite{Jensen2015,jensen2017strain}. 

In view of the above state of art, we chose crosslinked polyacrylamide hydrogel as the material with which to study the elastowetting phenomenon. One could easily modulate the associated shear modulus by varying the monomer concentration, while the gel-air surface tension $\gamma_{\textrm{G}}$ always remained close to that of water in air~\cite{Chakrabarti2014eml}. We explored the arrested wetting behavior of soft hydrogel spheres when placed upon smooth silanized silicon wafers of varying surface energies, as characterized through the work of adhesion $W$. In all cases, the elastocapillary length $\gamma_{\textrm{G}}/\mu$ and the elastoadhesive length $W/\mu$ were in the sub-millimetre range, so that solid-capillary and adhesion effects could be directly visualized through optical microscopy techniques. Moreover, the radii of the gel spheres were chosen to be smaller than their capillary length to avoid the effect of gravity. Our main objective was to measure the equilibrium contact angle $\theta^*$ of the soft elastic spheres on the various substrates, as a function of the gel shear modulus $\mu$, sphere radius $R$, and work of adhesion $W$, in order to quantify how they deviate from the pure wetting case. During our investigations, we found the presence of a localized contact foot of length $\ell$ in the arrested wetting configuration of these gel spheres. This is in essence the scenario envisioned by Joanny \etal~\cite{Joanny2001} but with different geometry and material. To rationalize our experimental observations, we develop a minimal analysis involving energy minimization, volume conservation, and scaling arguments.

\section{Experiments}
\begin{figure}[!t]
\begin{center}
\includegraphics[width=0.9\textwidth]{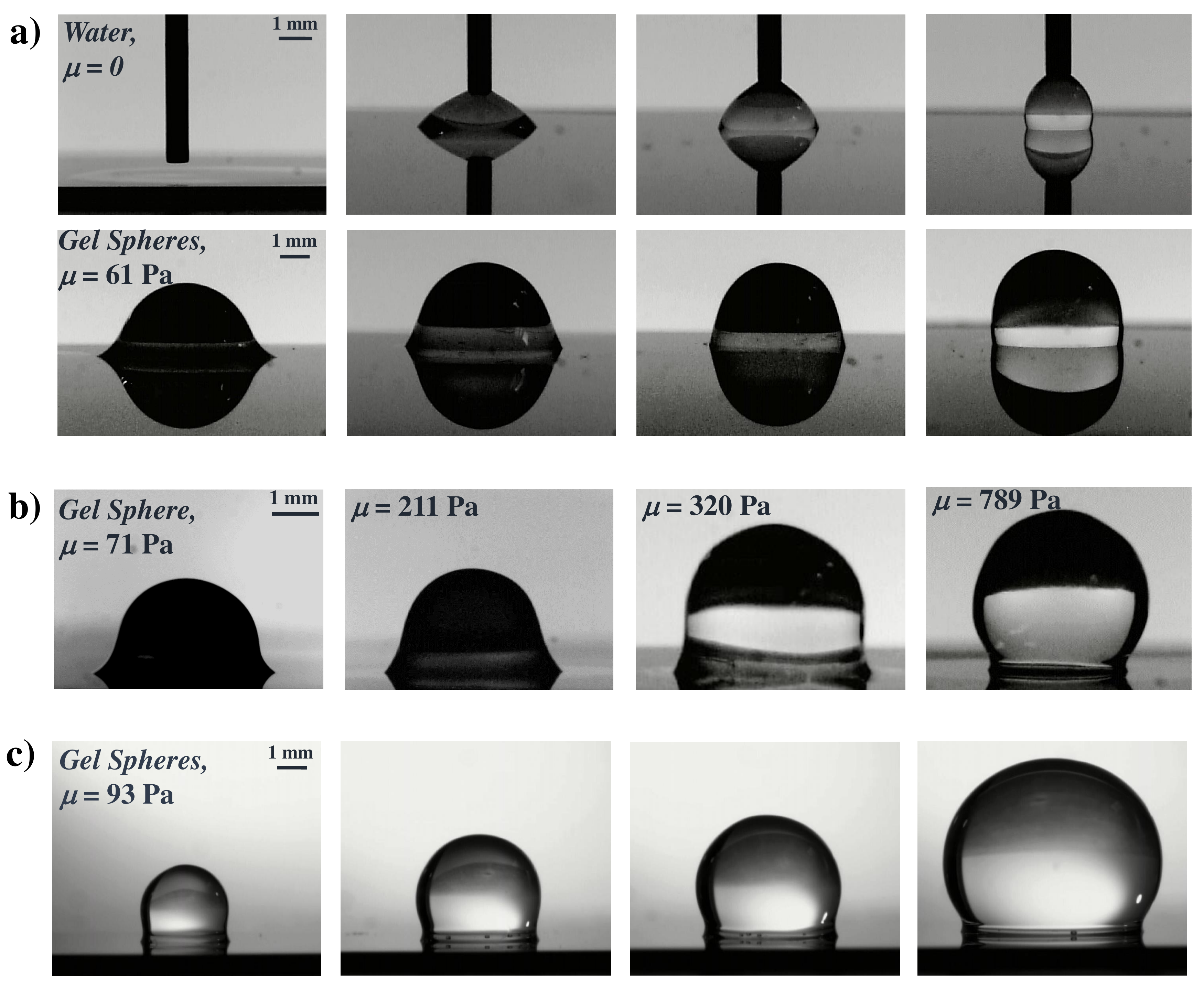}
\end{center}
\caption{(a) Comparison between equilibrium profiles of water drops (top) and gel spheres (shear modulus $\mu$= 61 Pa) (bottom) on silanized silicon wafers, for various works of adhesion $W$ (decreasing from left to right). (b) Equilibrium profiles of gel spheres on cleaned untreated silicon wafers, for various shear moduli $\mu$ (increasing from left to right). (c) Equilibrium profiles of gel spheres ($\mu$=93 Pa) on silicon wafers where water contact angle is 60$^o$, for various sphere radii $R$ (increasing from left to right). Scale bars of 1 mm depict calibrations for each row.} 
\label{fig : CAadhesion}
\end{figure}

In order to synthesize spherical hydrogel beads of different shear moduli~\cite{Chakrabarti2014eml}, we first prepared pre-gel solutions by diluting varying concentrations ($<$10\%) of the monomer N-(hydroxymethyl)-acrylamide (48\% solution in water, Sigma Aldrich) in deionized water. Upon degassing the monomer solutions by purging them with bubbles of pure nitrogen gas, we dissolved potassium persulphate (99.99\% trace metals basis, Sigma Aldrich; 0.25\% weight basis) and N,N,N',N'-tetramethylethylenediamine (TEMED, $\geq$99.5\%, purified by re-distillation, Sigma Aldrich; 0.3\% weight basis) into the solution that would initiate the polymerization of the polyacrylamide hydrogel roughly ten minutes after mixing the TEMED. The hydrogel spheres were prepared by suspending drops of the pre-gel solution in a beaker containing a vertical liquid density gradient created by introducing a heavy silicone oil (poly[dimethylsiloxane-co-methylphenylsiloxane], Sigma Aldrich, density $\rho = 1.05$~g/cc) at the bottom of the beaker and a lighter $n$-octane (99\% pure, Acros Organics, $\rho = 0.71$~g/cc) at the top gently using a pipette. The interface between these two oils was allowed to be well diffused by waiting for about thirty minutes to attain a uniform composition in the mixed zone. The ratio of the volumes of silicone oil to octane used in these experiments was around 2:1. Drops of pre-gel solution of different volumes were released gently into the octane layer. They sank to the diffused interface of octane/silicone oil until they were neutrally buoyant. Each pre-gel drop assumed a spherical shape at the interface and care was taken to allow enough space in between each drop to avoid coalescence. Even though the two liquids used in creating the density gradient had slightly different surface tensions with water, the broad thickness of the diffused zone ensured uniform curvature of the spheres. The drops of pre-gel solution cured into spherical elastic beads in about two hours at room temperature. As was reported earlier~\cite{Chakrabarti2014eml}, when these hydrogel spheres are placed atop a rigid substrate subjected to random vibration in the vertical direction, they exhibit surface harmonic modes which marginally differ from those of pure water drops. For liquids, surface modes are related to the surface tension and mass of the drops. For the surface modes of gels, the same scaling with surface tension and mass was observed, while the slight deviation of the values could be accounted for by the effect of elasticity. It was concluded that the surface tension of hydrogels with air is similar to that of water with air. The elastic shear modulus of these gels was measured ($\mu$ ranging from 61~Pa €to 789~Pa) using an oscillatory shear rheology technique that is described in more detail in the same previous study~\cite{Chakrabarti2014eml}. While preparing a batch of gel spheres of a particular monomer concentration, we cured a small amount of the same pre-gel solution into a rectangular slab of gel confined between two rigid plates, which was then set to horizontal vibration and the shear modulus was estimated from the slab's resonant mode.    

For preparing the rigid substrates, silicon wafers were cut into small pieces of about $2\ \textrm{cm}^2$ each and were flame-treated to remove all organic contaminants, rendering them hydrophilic. After cooling, they were placed in a chamber underneath a horizontal silane source, which rested upon spacers that were 13~mm tall. The silane source was prepared by attaching a flat sheet of filter paper to a glass slide with double-sided tape and depositing few drops of dodecyltrichlorosilane (Gelest Inc.) uniformly on the filter paper, wiping off any excess with tissue paper. The wafers were treated by the diffusing silane vapor in the chamber, at room temperature of 20$^\circ$ C and relative humidity of 35\% for different times. An exposure time of about 2 minutes in our system led to surfaces with a $\sim60^\circ$ contact angle for water, while an exposure time of about 12 minutes led to complete grafting of the surfaces giving a $\sim106^\circ$ contact angle for water -- the untreated surfaces being the most hydrophilic, where water spread as thin films. This resulted in the work of adhesion $W$, estimated from the Young-Dupr\'e equation for the different hydrogel-substrate systems, ranging from $\sim$50~mN/m to $\sim1$44~mN/m, corresponding to water contact angles of {$\sim106^o$} to {$\sim 0^o$} respectively (Fig. \ref{fig : CAadhesion}a, top). 

After curing, each hydrogel sphere was gently taken out from the liquid environment where they were prepared, using a plastic pipette with a smooth hydrophobic tip, and rinsed in pure $n$-heptane (Fisher Chemicals) repeatedly followed by moderate drying in air. The inner walls of the glass containers, which were used to house the heptane for cleaning, were also hydrophobized with vapors of dodecyltrichlorosilane to prevent the gel spheres from sticking to the walls thereby avoiding any possible damage to their surfaces. A treated silicon wafer was then placed on a weighing balance and the cleaned gel sphere was deposited upon it. This was the most critical step in our protocole as the pipette tip holding the gel sphere had to be held just above the wafer while slowly releasing the sphere on the latter to allow for uniform radial spreading. From the measured weight on the balance, the initial radius $R$ of the droplet was estimated (ranging from 1.2 mm to€" 3.8 mm) assuming a spherical shape. The gel-drop-on-substrate system was immediately photographed with a CCD camera (Sony XC-75). Each experimental measurement was completed in a few minutes, within which there was no observable loss in volume of the gel due to evaporation. On removing gel drops from hydrophilic substrates after a few minutes, we observed very thin and annular patches of liquid concentrated at the edge of the former contact region. However, the thickness of these annular films was always much smaller than the height of the foot region, thus ensuring that the measured macroscopic deformation of the gel-air interface, that gave us the modified contact angle $\theta^*$ and the foot length $\ell$, was not due to the traces of exuded liquid. These experiments being very sensitive to the smoothness of the gel surface, utmost care was taken while handling the spheres by pipette aspiration such that only a very small fraction was inside its tip. Furthermore, each gel sphere was used for the corresponding measurement only once as its surface got slightly damaged upon removing it from contact. 
\begin{figure}[t!]
\begin{center}
\includegraphics[width=1\textwidth]{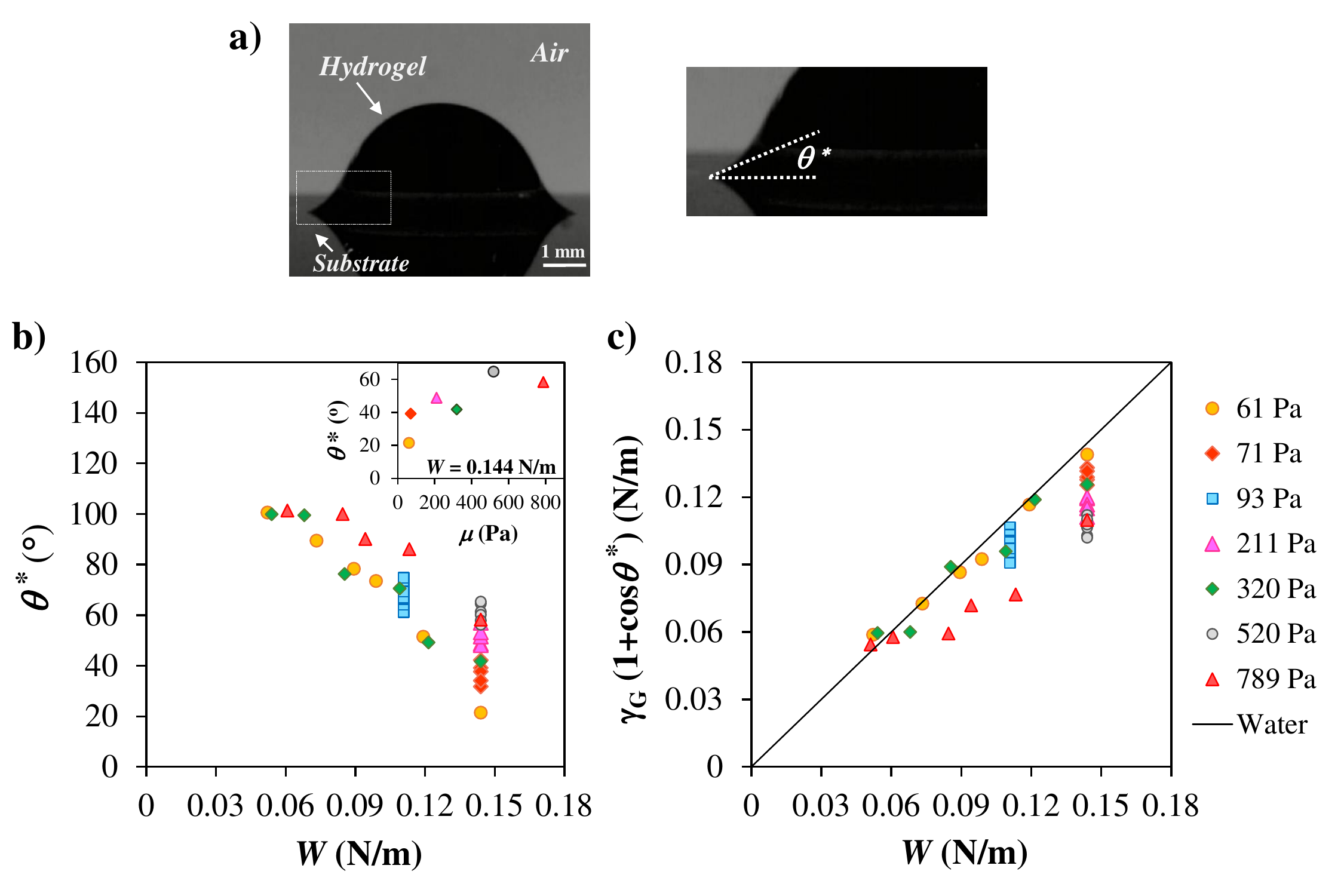}
\end{center}
\caption{(a) Snapshot of the equilibrium profile of a gel sphere ($\mu=$ 61~Pa) on a rigid substrate ($W=$ 144~mN/m). The highlighted region at the three-phase contact line is magnified on the right to show the measured contact angle $\theta^*$ that is obtained from a spline fitting of the entire gel-air interface~\cite{Stalder2006}. (b) Contact angle $\theta^*$ as a function of work of adhesion $W$ between the gel and the silanized silicon wafer in ambient air, for different shear moduli $\mu$ of the gel as indicated. (Inset) Contact angle as a function of shear modulus $\mu$ of the gel, for a substrate characterized by the work of adhesion $W=$ 144~mN/m. For comparison, on this substrate the contact angle of pure water is close to 0${^\circ}$. (c) Same data as depicted in (b), but plotted according to the Young-Dupr\'e law, where the contact angle $\theta^*$ is used, and where $\gamma_{\textrm{G}} \approx$ 72~mN/m in ambient air. The classical liquid reference is represented by the $y=x$ solid black line.}
\label{fig : CAAirData}
\end{figure} 

In sharp contrast with classical wetting of liquid droplets, the hydrogel spheres spread to their equilibrium configurations on the hydrophilic substrates by protruding a localized axisymmetric €œfoot€ at the edge of the contact region, while maintaining their undeformed spherical shape far away from contact (Fig.~\ref{fig : CAadhesion}a). Note that there was no such foot observed on the hydrophobic substrates. We estimated the length $\ell$ of the foot from the 2D side view in the following way (Fig.~\ref{fig : LipScalingData}a). A circle with the radius $R$ of the undeformed sphere was fitted to the upper periphery of the deformed gel sphere. A horizontal line along the reflection plane at the rigid substrate was drawn by joining the left and right triple-phase contact points (2D projection of the triple-phase contact line). The distance between one of these two triple-phase contact points and the nearest intersection of the fitted circle with the horizontal line defines the foot length $\ell$, and we used the averaged value from both the left and right sides. To measure the contact angle, the entire deformed gel-air interface was fitted with a spline curve, in ImageJ via the DropSnake~\cite{Stalder2006} plugin. An automatic algorithm then determined the contact angle $\theta^*$ at the triple-phase contact points (Fig.~\ref{fig : CAAirData}a), and we used the average value from both the left and right sides.

The contact angle $\theta^*$ of the hydrogel droplets on the silanized wafers decreases with increasing work of adhesion $W$ (Figs.~\ref{fig : CAadhesion}a and~\ref{fig : CAAirData}b), as for pure water, with a general trend reminiscent of the Young-Dupr\'e law (Fig.~\ref{fig : CAAirData}c). However, interestingly, $\theta^*$ increases with the shear modulus $\mu$ of the gel (Figs.~\ref{fig : CAadhesion}b and~\ref{fig : CAAirData}b inset), while being mostly independent of the droplet radius $R$ within our experimental range which was rather narrow by design in order to stay below the capillary length. Besides, as $\mu$ increases, the deformation and thus the foot length $\ell$ decrease (Fig.~\ref{fig : CAadhesion}b). The observation of a localized contact foot is the first verification of the prediction by Joanny \etal ~\cite{Joanny2001}. Nevertheless, our systematic study with gels involving different elastic moduli and radii, as well as substrates with varying wetting properties, reveals an even richer scenario. The latter will become more obvious below, as we attempt to recast the main experimental features within a theoretical framework involving energy minimization, volume conservation, and scaling arguments.

\section{Discussion}
The equilibrium profiles of hydrogel spheres placed on hydrophilic substrates suggest that elasticity, surface tension and adhesion are intimately coupled, and that there are two essential observables: the modified contact angle $\theta^*$, and the foot length $\ell$. In the following, we discuss those two quantities successively with minimal arguments. Specifically: for the contact angle, we consider an energy functional minimization and focus on the boundary term; while for the foot length, we avoid the detailed analysis of the resulting Euler-Lagrange equation and instead resort to simple scaling arguments. 

\subsection{Contact angle}
\begin{figure}[h]
\centering
\includegraphics[width=0.7\textwidth,keepaspectratio]{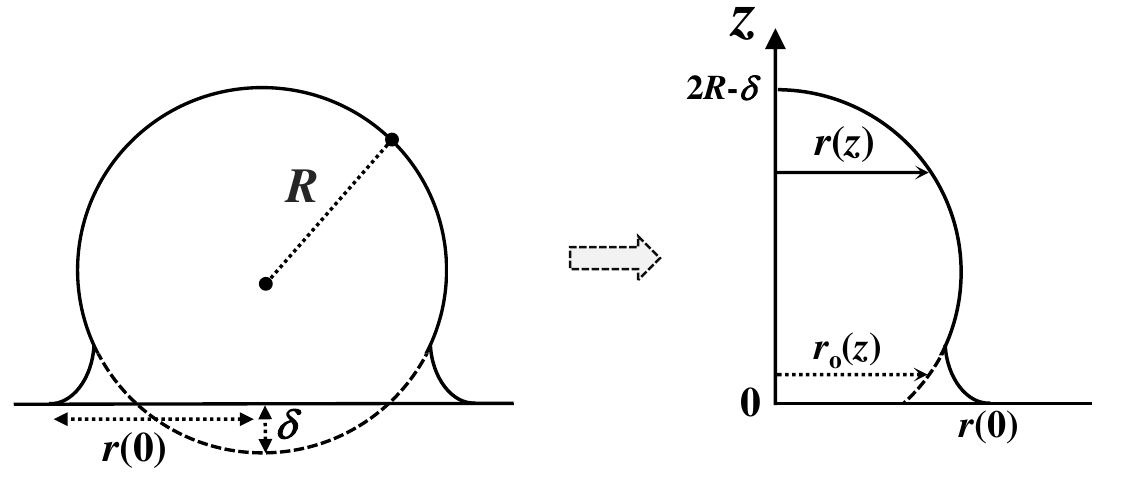}
\caption{Schematics of the system. A gel sphere with initial radius $R$ undergoes an axissymmetric deformation when placed on a rigid hydrophilic substrate ($z=0$), and $r(z)$ describes the radial distance of the gel-air interface in that deformed state as a function of $z$. As a reference, we introduce the radial distance $r_0(z)$ of the gel-air interface in the undeformed spherical shape (dashed circle). The entire contact radius (including the foot) is denoted by $r(0)$, and the vertical indentation depth by $\delta$.}
\label{fig:geo}
\end{figure}

The system and notations are defined in Fig.~\ref{fig:geo}.
Removing the constant bare-substrate surface energy, and including volume conservation, we consider the following energy functional:
\begin{equation} \label{eq1}
\begin{split}
 E\left[r(z)\right] =&(\gamma_{\textrm{SG}}-\gamma_{\textrm{SV}})\pi r(0)^2 +2\pi \gamma_{\textrm{G}} \int_{0}^{2R-\delta}\textrm{d}z\,r\sqrt{1+r'^2}+\lambda\left (\frac{4\pi R^3}{3}-\int_{0}^{2R-\delta}\textrm{d}z\,\pi r^2\right)\\ &+\frac{3\pi\mu}{2} \int_{0}^{2R-\delta}\textrm{d}z\,\left(r^2-r_0^{\,2}\right)\left(\frac{r-r_0}{r_0}\right)^2+\frac{\pi \mu}{2} \int_{0}^{2R-\delta}\textrm{d}z\, \left(r^2-r_0^{\,2}\right)\left(r'-r'_0\right)^2\ ,
  \end{split}
\end{equation} 
where the prime indicates the derivative with respect to $z$. The first term describes the energy of the gel-substrate interface, where $\gamma_{\textrm{SG}}$ and $\gamma_{\textrm{SV}}$ denote the substrate-gel and substrate-vapor interfacial tensions, respectively. The second term describes the energy of the gel-vapor interface. The third term ensures volume conservation through a Lagrange multiplier $\lambda$ with the dimension of a stress. The last two terms are \textit{ad-hoc} approximations for the elongational and shear elastic contributions, respectively, that assume incompressibility and enforce the strain to be localized in the foot region. Minimizing the energy functional with respect to the function $r(z)$, one obtains a Euler-Lagrange equation as well as two boundary conditions, one at $z=2R-\delta$ and one at $z=0$. We only consider the latter, that reads:
\begin{equation}
\cos\theta^*=\frac{\gamma_{\textrm{SV}}-\gamma_{\textrm{SG}}}{\gamma_{\textrm{G}}}-\frac{\mu}{2\gamma_{\textrm{G}}} \frac{r(0)^2-a^2}{r(0)}\left[r'_0(0)-r'(0)\right]\ ,
    \label{eqn : eq2}
\end{equation} 
with the notation $a=r_0(0)$ (Fig.~\ref{fig : LipScalingData}a), and where we introduced the contact angle $\theta^*$ (Fig.~\ref{fig : LipScalingData}b) geometrically defined by $\cos\theta^*=-r'(0)/\sqrt{1+r'(0)^2}$. In absence of elasticity ($\mu=0$), this condition reduces to the Young-Dupr\'e law: 
\begin{equation} 
\cos\theta=\frac{\gamma_{\textrm{SV}}-\gamma_{\textrm{SG}}}{\gamma_{\textrm{G}} }\ ,
\end{equation}
where $\theta^*=\theta$ is the classical contact angle of the liquid case. In contrast, when $\mu \neq 0$, $\theta^*$ is impacted by elasticity and thus different from $\theta$, and the boundary condition can be approximated by:
\begin{equation} 
  \cos \theta -\cos \theta^* \simeq \frac{\mu \ell}{\gamma_{\textrm{G}}} \left(\cot \theta^* + \cot \alpha \right)\ ,
  \label{eqn : modContAngle}
\end{equation} 
at first order in $\ell/a$, with the foot length defined as $\ell=r(0)-a$ (Fig.~\ref{fig : LipScalingData}a), and where we introduced the inner angle $\alpha$ of the foot region (Fig.~\ref{fig : LipScalingData}b). 

\begin{figure}[!h]
\begin{center}
\includegraphics[width=0.6\textwidth]{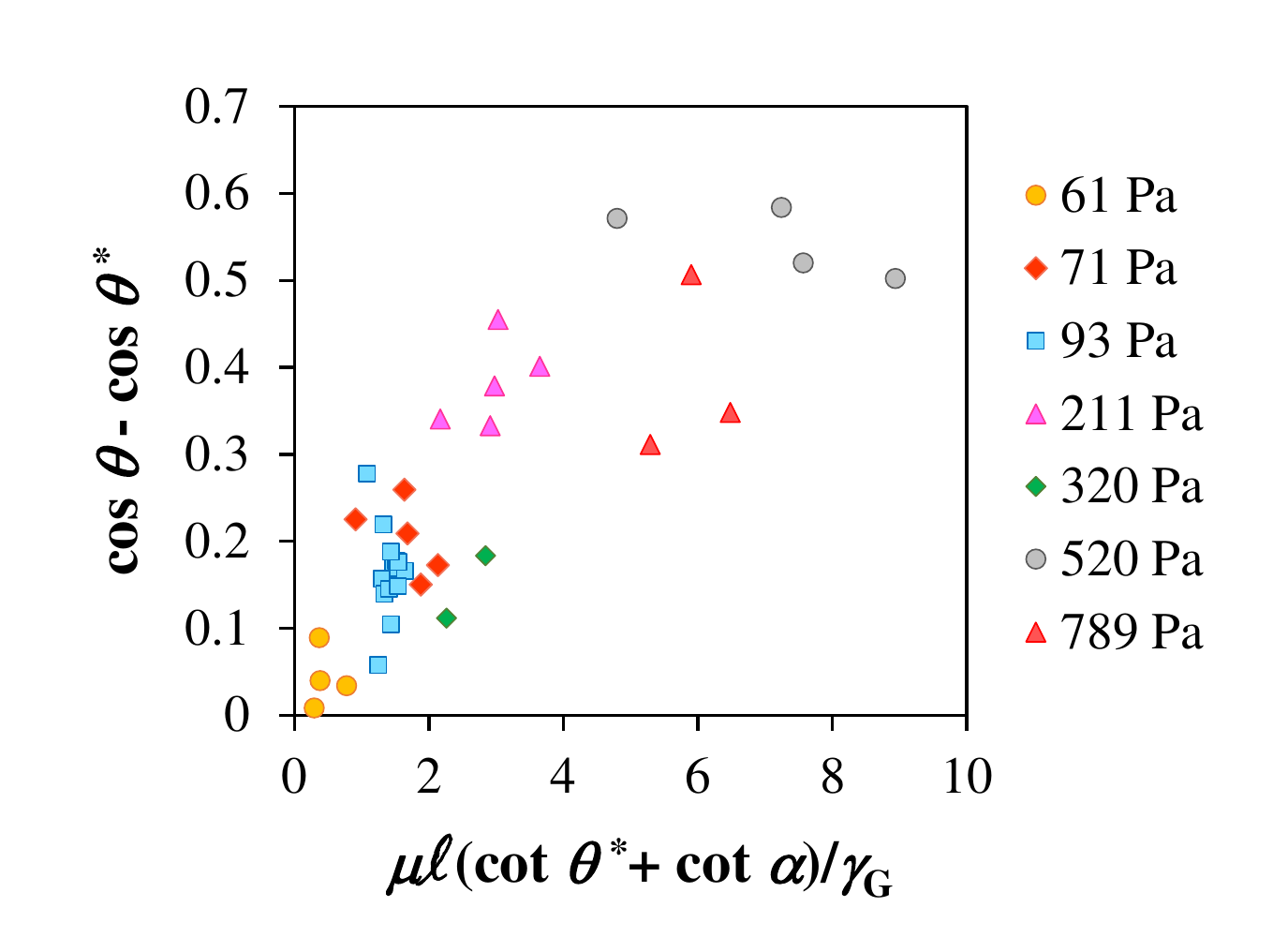}
\end{center}
\caption{Experimentally measured difference between the cosines of the contact angle ($\theta$) of water and the contact angle ($\theta^*$) of hydrogel, plotted according to Eq.~\ref{eqn : modContAngle}. The shear moduli are indicated in legend.}
\label{fig : theoryScaling}
\end{figure}
Figure~\ref{fig : theoryScaling} shows that the general trend of the experimental data is consistent with Eq.~\ref{eqn : modContAngle}. The observed $\sim1/10$ prefactor might result from the fact that the experimental measurements provide only macroscopic angles instead of the actual microscopic ones; while the scattering of the data possibly indicates that a more realistic shear energy integral is needed. Nevertheless, an important outcome of the above analysis is that the contact angle $\theta^*$ depends on elasticity only through the shear deformation of the gel, but not through its elongational deformation.

\subsection{Foot length}
\begin{figure}[!t]
\begin{center}
\includegraphics[width=1\textwidth]{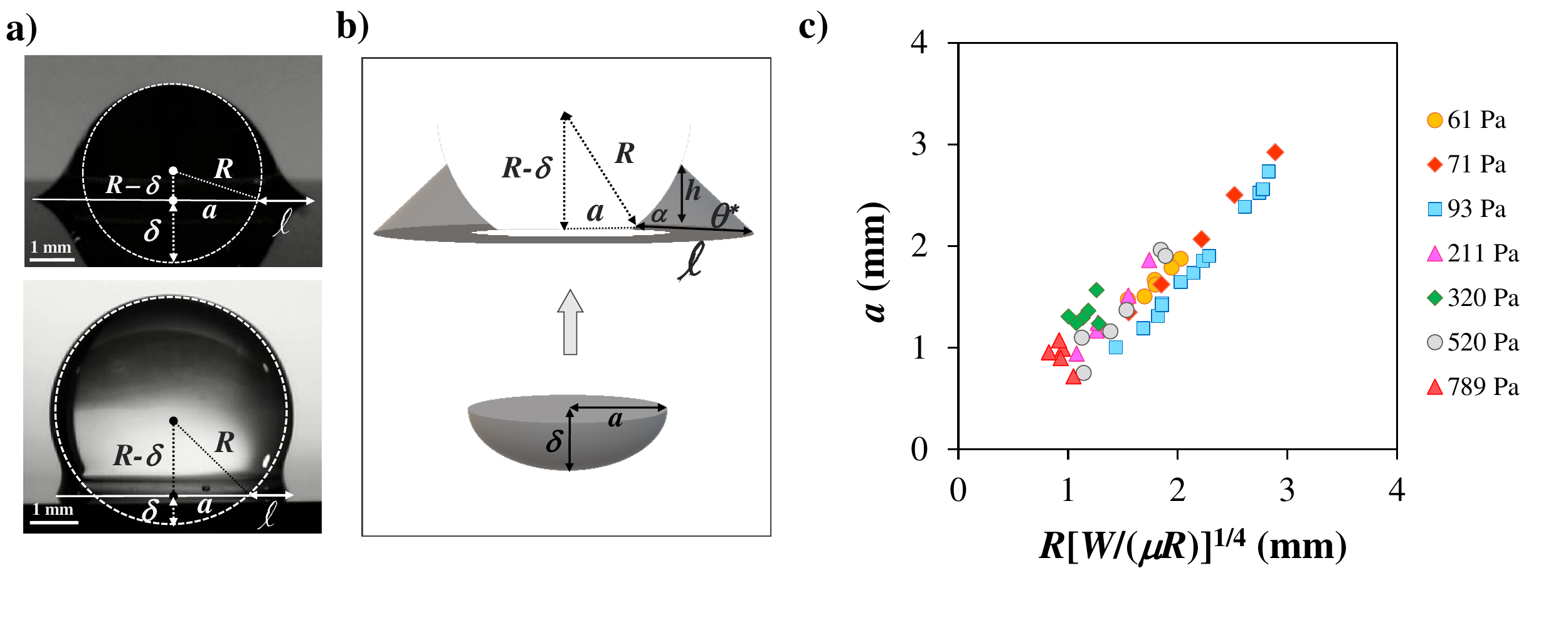}
\end{center}
\caption{(a) Two equilibrium profiles of gel spheres on hydrophilic substrates, one (top: $\mu=$ 61 Pa, $W=$ 0.144 N/m) with larger foot  length $\ell$, and the other (bottom: $\mu=$ 93 Pa, $W=$ 0.11 N/m) with smaller foot length. The undeformed spherical shapes (radius $R$) are depicted by the white dashed circles. The horizontal lines represent the reflection plane at the silicon wafer. The foot length $\ell$ is measured from the intersection of the dashed circle with the horizontal line to the triple contact line, while $\delta$ denotes the indentation depth (\textit{i.e.} the depth of the dashed circle underneath the horizontal line), and $a$ denotes the inner contact radius (\textit{i.e.} half the length of the horizontal line that is inside the dashed circle). (b) 3D schematic highlighting the volume conservation (grey) during the deformation of the incompressible gel: the volume of the fictive spherical cap below the substrate, with depth $\delta$ and base radius $a$, is redistributed in the axissymmetric triangular-like foot, with base $\ell$, height $h$, outer contact angle $\theta^*$ and inner angle $\alpha$. (c) Experimentally measured inner contact radius $a$ (Fig.~\ref{fig : LipScalingData}a) as a function of a combination of the relevant physical parameters, according to Eq.~\ref{inner}. The shear moduli are indicated in the legend.} 
\label{fig : LipScalingData}
\end{figure}
\begin{figure}[h]
\centering
\includegraphics[width=0.9\textwidth,keepaspectratio]{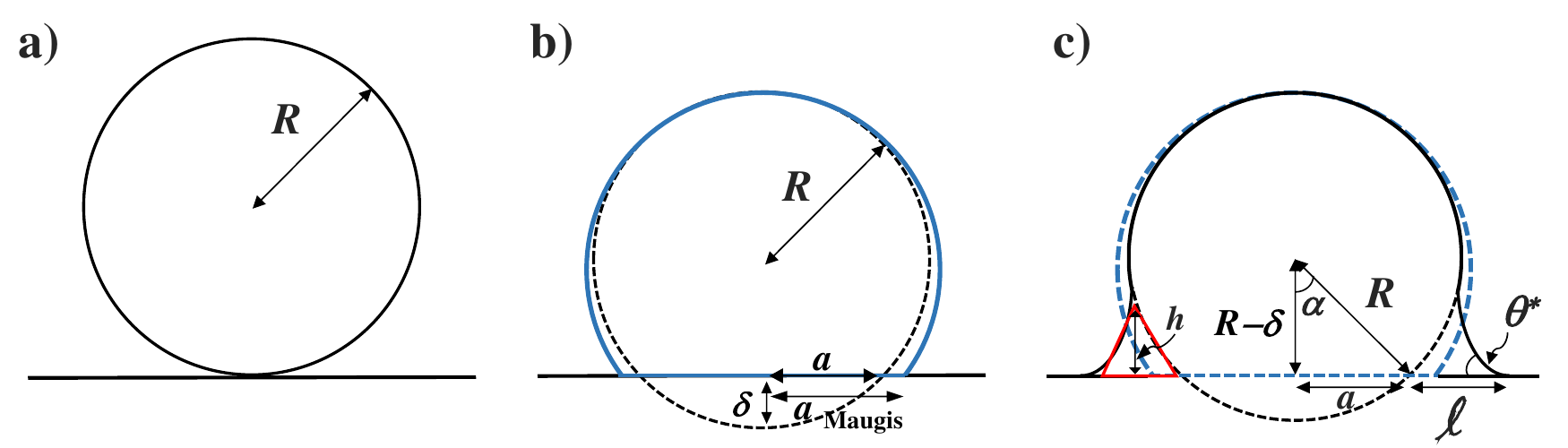}
\caption{(a) The gel sphere, with radius $R$, just before contact with the flat rigid substrate below. (b) In the first step of the decomposed elastowetting process, the gel undergoes a pure Maugis elastic deformation, \textit{i.e.} a geometrically-large-strain JKR-like deformation~\cite{Johnson1971,Johnson1985}, thus essentially creating a spherical cap with a base radius $a_{\textrm{Maugis}}$. (c) In the second step, the local capillary action near the triple-phase contact line further extracts an axissymmetric triangular-like foot, at constant volume and inner contact radius (or equivalently indentation depth $\delta$).}
\label{fig : jkrTOours}
\end{figure}

The  superposition of the image of the gel in the undeformed spherical state and that resulting from the wetting-induced deformation (Fig. \ref{fig : LipScalingData}a) shows that the upper portion of the gel remains essentially undeformed, whereas the lower region undergoes a foot-like deformation. Such an observation deviates from classical contact mechanics of elastic objects, in which the elastic sphere essentially maintains a spherical-cap shape (Fig. \ref{fig : jkrTOours}b)~\cite{Johnson1971,Johnson1985}. The base radius $a_{\textrm{Maugis}}$ of the latter was analyzed by Maugis~\cite{Maugis1995} within the framework of JKR theory~\cite{Johnson1971,Johnson1985}, and found to scale as $a_{\textrm{Maugis}}\sim R\left[W/(\mu R\right)]^{m}$, with $m$ being close to $1/4$ at geometrically large deformations~\cite{rimai1989adhesion}. Volume conservation due to the  incompressibility of the elastic material also imposes that $a_{\textrm{Maugis}}$ is geometrically related to the inner contact radius $a$ of the (shifted) undeformed sphere (Fig. \ref{fig : jkrTOours}c), as:
\begin{equation}
\frac{a_{\textrm{Maugis}}}{R}=\sqrt{\frac{8-{\left[1+\sqrt{1-\left(\frac{a}{R}\right)^2}\right]}^3}{3\left[1+\sqrt{1-\left(\frac{a}{R}\right)^2}\right]}}\ ,
\label{eqn : aWithBeta}
\end{equation}

We now focus on the situation where $0<a/R<0.9$, which corresponds to the range of our experiments, and which already contains geometrically large deformations. In that situation, the error one makes by replacing the right-hand side of Eq.~\ref{eqn : aWithBeta} by $a/R$ is inferior to $20\,\%$, which is acceptable given the wide range of parameters studied in the experiments. Therefore, combining the above ingredients, one gets the following reasonable approximation:
\begin{equation}
a\approx R\left(\frac{W}{\mu R}\right)^{1/4}\ .
\label{inner}
\end{equation}
While we do not expect the real contact radius $r(0)$ of our experiments to be given by $a_{\textrm{Maugis}}$, as already explained above, that approximation for the inner contact radius $a$ is valid, as shown in Fig. \ref{fig : LipScalingData}c, with the missing prefactor being close to unity.

As a consequence of the previous analysis, we are led to propose the following elastowetting scenario with a schematic decomposition in two steps: first the sphere undergoes a Maugis-like elastic deformation (Fig.~\ref{fig : jkrTOours}b) inducing a spherical-cap shape characterized by its contact radius $a_{\textrm{Maugis}}$ and the inner contact radius $a$ (or equivalently the indentation depth $\delta$); secondly, from that intermediate reference stage, the local capillary action near the triple-phase contact line extracts a foot, but at constant volume and inner contact radius (or equivalently indentation depth $\delta$), leading to a reorganization of the shape near contact (Fig.~\ref{fig : jkrTOours}c).
\begin{figure}[!t]
\begin{center}
\includegraphics[width=0.5\textwidth]{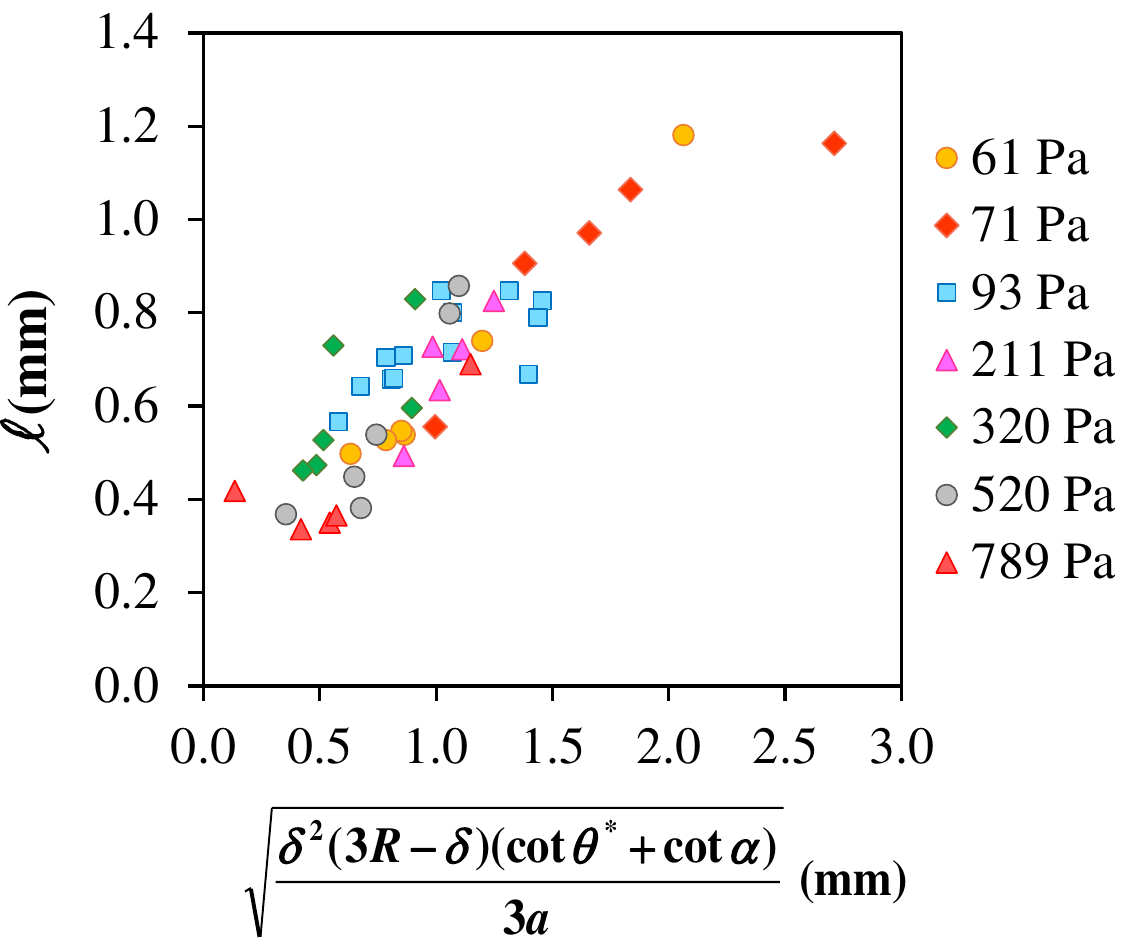}
\end{center}
\caption{Experimentally measured foot length $\ell$ as a function of a combination of the relevant geometrical parameters, according to Eq.~\ref{eqn : lip}. The shear moduli are indicated in the legend.}
\label{fig : LipscalingGraph} 
\end{figure}
To further characterize the foot length $\ell$, we balance the volume $\pi \delta^2(3R-\delta)/3$ of the fictive spherical cap below the substrate (Fig. \ref{fig : LipScalingData}b), and the volume $\sim\pi a \ell h$ of the axisymmetric foot at lowest order in $\ell/a$. Indeed, to a fair approximation, the cross section of the latter can be considered to be triangular, with base $\ell$ and height $h$, such that $\ell = h \left(\cot \theta^*+\cot \alpha \right)$. Combining those ingredients, we obtain:
\begin{equation}
 \ell \approx \sqrt{\frac{\delta^2(3R-\delta)(\cot \theta^* + \cot \alpha)}{3a}}\ .
\label{eqn : lip}
\end{equation}
This relation agrees well with experiments, as shown in Fig.~\ref{fig : LipscalingGraph}, with the missing prefactor being close to unity.

As a final remark, while two central relations (Eqs.~\ref{eqn : modContAngle} and~\ref{eqn : lip}) have been obtained and checked, they in fact involve five constitutive variables: $\theta^*$, $\ell$, $\alpha$, $a$ and $\delta$. Thus, three other independent relations are needed to relate each of these variables only to the physical parameters of the problem: the initial radius $R$ of the sphere, the shear modulus $\mu$ of the gel, the work of adhesion $W$ between the gel and the substrate in air, and the gel-air surface tension $\gamma_{\textrm{G}}$. Those three other relations are: Eq.~\ref{inner} (Fig.~\ref{fig : LipScalingData}c), $\delta=R-\sqrt{R^2-a^2}$ (Fig.~\ref{fig : LipScalingData}a), and $\cot \alpha = (R-\delta)/a$ (Fig.~\ref{fig : LipScalingData}b), thus bringing closure to the problem.

\section{Conclusion}
We presented novel experimental results on how soft hydrogel spheres deform when in contact with rigid substrates of varying surface energies. On hydrophilic surfaces, these spheres protrude an axisymmetric foot in the contact region, while maintaining a global undeformed spherical shape. Furthermore, while the contact angle of these gel spheres decreases with increasing substrate'€™s wettability, as expected, it is found to increase with the elastic modulus of the gel. We rationalized these observations with a minimal analysis based on energy minimization, volume conservation, and scaling arguments. In future work, nonlinearities ensuing from both materially and geometrically large deformations of the gel should be incorporated within the framework of the Euler-Lagrangian equation introduced here. Along with this possible theoretical refinement, more precise experiments, with \textit{e.g.} confocal microscopic techniques~\cite{Style2013}, would as well be of paramount importance.  

Among possible applications, one could use the finite contact angle of soft gel spheres on hydrophilic substrates to estimate the surface energies of those substrates, when liquids usually spread as thin films debarring them as suitable candidates for that task. Additionally, our results on elasticity-dependent contact angles of gels could be applied to characterize the viscoelastic foot that is observed in the peeling of pressure-sensitive adhesives from a substrate. While the role of the shear stress in the contact angle and length of such viscoelastic foots was implied in previous studies~\cite{newby1995science,newby1997langmuir}, it is our hope that the current work will bring more attention to such a matter. Last but not least, the associated dynamic processes, involving friction and the possible slippage of those viscoelastic foots, represent important extensions of the present problem.

\section{Acknowledgments}
The authors thank Basile Audoly and Animangsu Ghatak for helpful discussions. A.P. was financially supported by LabEx ENS-ICFP: ANR-10-LABX-0010/ANR-10-IDEX0001-02 PSL. T.S. acknowledges funding from the Global Station for Soft Matter, a project of Global Institution for Collaborative Research and Education at Hokkaido University, as well as Howard Stone for hosting him in his group during part of this project. M.K.C. thanks the Chair Total - ESPCI Paris for financial support.

\bibliography{Chakrabarti2017.bib}
\end{document}